# Solar Mass Loss, the Astronomical Unit, and the Scale of the Solar System

Peter D. Noerdlinger[*]

*St. Mary's University, Department of Astronomy and Physics*

*Halifax, N.S., B3H 3C3, Canada*

e-mail: pnoerdlinger@ap.stmarys.ca

The radiative and particulate loss of mass by the Sun, $\dot{M}_\odot / M_\odot = -9.13 \times 10^{-14} \, y^{-1}$ or more causes the orbits of the planets to expand at the same rate, and their periods to lengthen at twice this rate. Unfortunately, under the present definition of the Astronomical Unit (AU) based on the fixed Gaussian gravity constant $k_{GS} = 0.01720209895 \, (AU)^{1.5} \, day^{-1}$, the value $AU_{met}$ of the AU in meters must *decrease* at 1/3 this rate, all these rates being expressed logarithmically. The progress of the planets along their orbits slows quadratically with time. For example, in one century Mercury would lag behind the position predicted using constant solar mass by almost 1.4 km, in two centuries 5.5 km. The value of $AU_{met}$ can be made constant by redefining it, based on a *reference* solar mass unit, such as $M_{\odot,J2000}$; else, the solar Gaussian gravity constant $k_{GS}$ used in defining the AU could be redefined $\propto \sqrt{M_\odot}$. Improved accuracy of the ephemerides would impose useful bounds on losses due to axion emission (Sikivie 2005). With no axion emission the Earth's semi-major axis grows 1.37 m/cy; with the maximum allowable such emission the result is 1.57 m/cy. Under reasonable assumptions about alternate gravity theories, radar delay data are used to show that the effect of a changing Newtonian gravity constant is negligible.

---

[*] Present address: 1222 Oakleaf Circle, Boulder, CO 80304

## 1. Introduction

As astrometry improves towards the micro arc second level, and as clocks near a long term stability level of parts in $10^{18}$ per year (Tanaka et al 2003, Oskay et al 2006)[1], smaller and subtler perturbations will be detectable with conventional and radar astrometry in the solar system. Most studies to date incorporate many perturbations to planetary motion, and allow for the inclusion of more such terms as measurements are refined. The secular effect of the Sun's loss of mass was traditionally omitted (Guinot 1989, Newhall et al 1983, Standish 1995) or given only brief mention (Brumberg 1991), but has recently been studied (Krasinsky and Brumberg 2004, Pitjeva 2005a, 2005b). Krasinsky and Brumberg (2004) (hereinafter KB), conducted both theoretical and observation-based analyses but omitted the dominant contributor to mass loss, namely radiation, including only a somewhat overestimated loss to the solar wind. By contrast, Pitjeva's work is based purely observational data and analysis thereof, so it is insensitive to mass loss mechanism. Simulations of planetary motion that range from high accuracy ephemerides spanning 44 to 60 centuries (Newhall et al 1989, Standish et al 2005) to even longer term computations investigating chaos (Laskar 1994, Wisdom and Holman 1991, Ito and Tanikawa 2002, Hayes 2007), are already sensitive at the level where solar mass loss should be considered. Duncan and Lissauer (1998) have analyzed stability in the post-main-sequence regime, and find gross effects

---

[1] Pulsar timing data can be used to supplement, or possibly some day to supplant, the use of atomic clocks (Hobbs et al 2006)

due to the larger mass loss rate expected then. In the sequel we shall see that, due to complications in the definition of the AU, the rate of change of planetary major axes is *not* the same as that of the AU itself; in fact, they are of *opposite sign*. By and large, we shall assume that, as in common parlance, an estimated rate of change of the AU quoted here refers to that of the Earth's semi-major axis $a_\oplus$, in meters, with the orbits of the other planets scaling in proportion. In summary, we consider physically observable effects (Section 6), as well as problems of definitions (Section 5).

## 2. Contributors to solar mass loss and their estimated values

The Sun has luminosity at least $L_{\odot,EM+\nu} = 3.939 \times 10^{26}\,\text{W}$, or $4.382 \times 10^9\,\text{kg}\,\text{s}^{-1}$ (Bahcall 1989), including electromagnetic ("EM") radiation $L_{\odot,EM} = 3.856 \times 10^{26}\,\text{W}$ and a ~2.3% contribution ("$\nu$") from neutrinos. Solar luminosity has varied less than 0.1% in the last 2 to 3 centuries (Sofia and Li, 2000). Most of the variation would presumably be due to changes in surface activity, and would not be secular. The particle mass loss rate (solar wind) is about $1.374 \times 10^9\,\text{kg}\,\text{s}^{-1}$ (Hundhausen 1997). There could also be a loss by radiation of axions (Raffelt 1999, Sikivie 2005) amounting to as much as 10% or more of the electromagnetic luminosity. An approximate but compelling upper limit for axion losses of $L_{axion} < 0.2 * L_{\odot,EM}$, or 20% of the EM part, is found by helioseismology (Schlattl et al, 1999). Large values might also impose unacceptable changes on the established agreement of main sequence stellar evolution models and main sequence Herzsprung Russell diagrams. Let us define the ratio of axion losses to electromagnetic as

$$axratio = L_{axion} / L_{\odot,EM} \; ; 0 \leq axratio < 0.2 \qquad (2.1)$$

and express all related results either in terms of *axratio*, or as a range, it being understood that the lower end of the range is for no axion losses, the upper for *axratio* = 0.2 . The mean rate of accretion of solid matter onto the Sun is negligible (Brownlee, 1996). The solar mass loss rate due to EM radiation, neutrino losses, and the solar wind is then,

$$\dot{M}_{\odot,EM,\nu,wind} = -5.75 \times 10^9 kg \; s^{-1} = -9.13 \times 10^{-14} M_\odot \; y^{-1}, \qquad (2.2)$$

but these values could be enhanced by axion emission, as expressed by

$$\dot{M}_{\odot,EM,\nu,wind,axion} / \dot{M}_{\odot,EM,\nu,wind} = 1 + 0.7442 * axratio \qquad (2.3)$$

The odd-looking coefficient 0.7442 is due to the expression of *axratio* as a fraction of the electromagnetic luminosity, while neutrinos and the solar wind make their own contribution. For comparison, KB used an older and larger estimate for the solar wind mass loss rate of $1.891 \times 10^9$ kg s$^{-1}$, and they did not take into account the radiative losses, which are more than double that number.

## 3. Consequences of the mass loss

Due to the losses just described, if *G* is the Newtonian gravity constant, the product $GM_\odot \approx 1.327 \times 10^{20} m^3 s^{-2}$ (McCarthy 1996, Standish 1994) steadily decreases. The main consequences are: (i) the orbits of the planets, minor planets, and short period comets will *expand*, at constant eccentricity, $\propto M_\odot^{-1}$ (ii) the Astronomical Unit (AU) (Standish 1994, McCarthy1996), as presently defined, will *shrink* $\propto M_\odot^{1/3}$, (iii) the periods of the planets will *increase* $\propto M_\odot^{-2}$ and (iv) the planets will lag, in their orbital motion, behind the positions predicted by theory with a static solar mass, the lag being proportional to the square of the elapsed time. I shall omit discussing the gravitational effect of the solar radiation and solar wind. The gravitational effect of the solar wind is very small, and has negligible secular effect on the major axes, while that of the radiation would have to be treated through a relativistic discussion. Newtonian theory suffices to estimate the small effects discussed here. Note that although it is sometimes said (Klioner 2008) that the effect of mass loss by the Sun is the same as that of a "changing" gravitational constant G, we consider the latter case as an error source, in Section 6.2.1. Mass loss from the Sun has negligible effect on lunar motion, but significant effects on planetary orbits, while a nonzero $\dot{G}/G$ affects both. Points (i) - (iv) are discussed in turn after some preliminary considerations on orbits.

## 4. Orbital analysis

Let $\ell$ denote the specific angular momentum of the planet, $\ell = r^2\omega$, a constant of the motion, and $\mu(t)$ denote the product *GM*; also let dots denote time derivatives, e.g. $\dot{\mu} \equiv d\mu/dt$.

The total energy (no longer conserved) is given by

$$E = -\mu/2a.  \qquad (4.1)$$

Note, further, that

$$\ell^2 = \mu a(1-e^2),  \qquad (4.2)$$

where *e* is the orbital eccentricity.

where *e* is the orbital eccentricity, and from Equation (3-43) of Goldstein (1950)

$$r^{-1} = \mu\left(1+e\cos\theta\right)/\ell^2. \tag{4.3}$$

Here $\theta$ is the "true anomaly" or angle round the orbit measured from perihelion, obeying

$$\dot{\theta} \equiv \omega = \ell/r^2 \tag{4.4}$$

The secular rates of change of *E*, *a*, *e*, $\omega$, and the period *P* are all of interest for solar system dynamics. The secular rates of change in *a* and *e* were derived by Jeans (1929), who obtained

$$\mu a = \mu_0 a_0 = constant;\ \ e = e_0 = constant \tag{4.5}$$

where the zero subscripts indicate the values at a reference time, say J2000. (Also see Weinberg, 1972). The same result was obtained by Kevorkian and Cole (1996) (hereinafter KC) using more rigorous methods. We accept their analysis as definitive. KC also found that to first order in $\dot{\mu}/\mu$ the rate of change of the argument of perihelion $\varpi$ is zero. To higher orders, KC found additional *periodic* terms in $\varpi$, but no *secular* terms. Other authors have found varying results, using other approximation methods. Deprit (1983) agrees with KC, but Hadjidemetriou (1966) found secular terms in *e* and in $\varpi$ for various assumptions (such as power laws) for slow time dependence of mass loss rate. KC showed in other examples (e.g. their § 1.3.1), however, that overly simple expansion methods can lead to spurious secular terms. Adopting the KC analysis, we find that the semi-major axis increases in inverse proportion to the mass, while the eccentricity and argument of perihelion are constant. Equations (4.2) and (4.5), and the constancy of the specific angular momentum $\ell$, also demonstrate, more simply, that the eccentricity $e$ is constant (Jeans, 1929). Differentiating Equation (4.1) yields $da/a = d\mu/\mu - dE/E$ so that, with Equation (4.5), $dE/E = 2d\mu/\mu$. The variation of the planetary period is found from Kepler's third law, which reads

$$\omega_{SI}^2 = \mu a^{-3}(1+m). \qquad (4.6)$$

where $m$ is the ratio of the sum of the planet's and its satellites' masses to $M_\odot$, and $\omega_{SI}$ is the angular orbital speed (we suppress the question of time units; see Guinot 1989, Klioner 2008). For example $m_\oplus \sim 1/328900$ (McCarthy 1996, Standish 1994). With Equation (4.5), this yields, neglecting henceforth the tiny variation of $(1+m)$ due to that in $\mu$,

$$\omega/\omega_0 = (\mu/\mu_0)^2. \qquad (4.7)$$

## 5. The fate of the Astronomical Unit

These physical results require re-examining the definition of the Astronomical Unit (AU). The definition is based on the value of $k_{GS}$, the solar Gaussian gravity constant. The International Astronomical Union (IAU) has designated $k_{GS}$ as a *defining constant* (McCarthy 1996, Standish 1994, Kaplan 2005), fixed at

$$k_{GS} \equiv 0.01720209895 \, (\text{AU})^{1.5} \, \text{day}^{-1} \qquad (5.1)$$

where the day is 86,400 seconds of TDB (barycentric dynamical time) (WS89; Standish, 1998). The AU is then defined as "the radius of a circular orbit, in which a body of negligible mass, and free of perturbations, would revolve around the Sun in $2\pi/k_{GS}$ days" (Blumberg and Boksenberg 1996). Its occurrence in Equation (5.1) is as a *unit*. If this definition and Equation (5.1) are retained, the value of the AU in m (denoted "$AU_{met}$")[2] will perforce undergo secular change as the Sun's mass decreases. [Note that, due to relativistic effects, the relation of TDB to TCB (SI time asymptotically far from the Sun) will slowly change as well, but only by parts in $10^8$ of the already small effects considered here.] To see why the AU would have to change, note that, according to the foregoing definition, the value of $AU_{met}$ can be determined from Kepler's Third Law expressed in AU and days, as in Equation (4) of Williams and Standish (1989) (hereinafter WS89):

$$\bar{\omega}_{R/d} = k_{GS} \left( a_\oplus / AU_{met} \right)^{-3/2} \sqrt{1+m} \qquad (5.2)$$

where $a = a_\oplus$ is the semi-major axis of the Earth's orbit in meters, $m$ its mass (with the Moon's) in solar units, and $\bar{\omega}_{R/d}$ its mean daily angular motion, expressed in radians per TDB day. In practice, WS89 fit the radar echo data to Mars' orbits in determining $AU_{met}$. (For Mars, $m \sim 1/3098708$) (McCarthy 1996, Standish 1994, Pitjeva 2005a). Determining the radar echo time fixes the scale of the semi-major axes $a$ in m, because $c$ is a defining constant. Given that $a$ and $\omega_{R/d}$ are measured, and that $k_{GS}$ is a defining constant, Equation (5.2) yields a unique value of $AU_{met}$ for any planet, and hence, by suitable averaging, an overall value that sets the scale of the Solar System. On combining the logarithmic derivatives of Equations (4.5), (5.1), and (5.2) *with the variation of $AU_{met}$ in Equation* (5.2) *included*, because it is a numerical conversion factor and not a unit, and with $k_{GS}$ as a *fixed numerical constant*, one finds that

$$AU_{met} / AU_{met,0} = (\mu / \mu_0)^{1/3} \tag{5.3}$$

---

[2] The somewhat awkward subscript is intended to skirt the question as to whether distances based on light transit times using different timescales are in SI or other units. It does not matter here so long as one is consistent.

In principle, then, the AU *decreases* as the cube root of the solar mass, while the semi-major axes of the planets *increase*! The planetary orbital radii, *in units of AU*, would have a net dependence on the Solar mass $\sim (\mu/\mu_0)^{-4/3}$, a rather arcane result seemingly unanticipated when the definitions of the AU and $k_{GS}$ were established. In practice, this situation could arise only if Equation (5.2) were applied to separate data sets representing different time periods within an overall long data set, an unlikely event, in view of the need to use all the available data. Yet, the situation is disturbing and invites a remedy; one does not want units to have time variations. At it stands, the changing AU impairs analyses of solar system geometry and dynamics, and, in principle, even corrupts the use of astronomical parallax as a distance measure. To ensure the constancy of the AU in meters, one has either to change the definition of the AU or to make $k_{GS}$ a function of $\mu$. The second option, which preserves the present definition of the AU, is considered first, although it would apparently be inconvenient for most solar system geodesists, who rely on the constancy of $k_{GS}$ in making units changes and in fitting the data to Equation (5.2). This option would require that Equation (5.1) be replaced by $k_{GS} = 0.01720209895 \, (\mu/\mu_0)^{1/2} \, (AU)^{1.5} \, day^{-1}$. $AU_{met}$ would then be constant, but its value would, with that of $k_{GS}$, be subject to determination of $\mu/\mu_0$, and would be sensitive to the choice of the base epoch defining $\mu_0$. This choice preserves the meaning of the Gaussian gravity constant as $\sqrt{GM_\odot}$. If one instead defined $k_{GS} = 0.01720209895 (\mu/\mu_0)^2 \, AU^{1.5} day^{-1}$, then $AU_{met}$ would scale like the semi-major axis $a$ of a planetary orbit, $AU_{met} \propto \mu_0/\mu$. This choice preserves concept of the AU as a natural scale for the Solar System.

A referee has suggested a different approach. If the number "1" in Equation (5.2) is replaced by the ratio $M_\odot / M_{\odot,ref}$, with the solar reference mass $M_{\odot,ref}$ (solar mass unit) fixed by some reference epoch, say J2000, this ratio will vary in such a way as to allow *both* $k_{GS}$ and $AU_{met}$ to be constant. The revised equation becomes

$$\bar{\omega}_{R/d} = k_{GS} \left( a / AU_{met} \right)^{-3/2} \sqrt{\left( M_\odot / M_{\odot,ref} \right) + m} \qquad . \tag{5.4}$$

The ratio $M_\odot / M_{\odot,ref}$ can be estimated through Equation (2.2), but axion losses are unknown and the solar wind loss may vary over extended time epochs, so the definition can apparently be implemented only with some uncertainty. This approach preserves the AU as a useful unit, however, in conjunction with setting a fixed Gaussian solar mass unit. *It involves, however, a redefinition of the AU.* The new definition would have to be: "The Astronomical Unit is defined as the radius of a circular orbit, in which a body of negligible mass, and free of perturbations, would revolve around a body whose mass is one solar mass unit in $2\pi/k_{GS}$ TDB days." The Gaussian solar mass unit would have to be tied to some epoch (Newhall et al 1983). One can regard Equation (5.2) as a variant of Equation (5.4) with $M_{\odot,ref}$ allowed, in fact, to vary, so that the solar mass unit is always the mass of the Sun. Using Equation (5.2), however, does not necessarily force the solar mass unit to be the current mass of the Sun. That interpretation it is quite workable, but all the present discussion except that which uses $M_{\odot,ref}$ could be conducted without reference to a solar mass unit. The use of Equation (5.2) is consistent with Kepler's Third Law and the current definition of the AU, without regard to mass units other than SI. Historically, defining $k_{GS}$ as a fixed number and ignoring $\dot{\mu}$ meant that any determination of $GM_\odot$ in SI units resulted in a redetermination of the AU. If Equation (5.4) and the new definition of the AU are adopted, with the solar mass unit $GM_\odot$ fixed to an epoch, the only change will be that we determine the AU by determining the SI value of $GM_\odot$. Klioner (2008) has suggested that the community might well just fix the AU in SI meters, based on some epoch, in which case it is just a unit of convenience, such as the kilometer or the mile. He suggests that the effects of solar mass loss and that of a changing constant of gravity are the same, though the former does not appreciably affect the lunar orbit, while the latter does.

## 6. Observational possibilities

### *6.1 Predicted effects*

#### 6.1.1 distances

To compare with experiment, the numbers of immediate interest are the variations of the semi-major axes of planetary orbits as compared with the estimated error in the AU, as well as secular corrections to the mean motion, namely changes in the period and mean or true anomaly. We present these results for a range of axion losses. For Earth we have, e.g.

$$da_\oplus / dt = +0.0137 \, \text{my}^{-1} \text{ to } +0.01574 \, \text{my}^{-1}; \tag{6.1}$$

for the range of axion loss from zero to 20% of the electromagnetic radiative losses.

In the 44 centuries spanned by the DE200 ephemeris (Newhall et al 1983), the increase in $a_\oplus$ amounts to 60 - 69 m, which is larger than their stated present error in the AU, 149,597,870,680 $\pm$ 30 m (Newhall et al, 1983). Within the timespan of accurate radar astrometry, ~ 46 y, the change is, however, immeasurably small. Even in the life of the Solar System, the total mass loss fraction is normally considered to be about 3.5 x $10^{-4}$, implying a negligible effect on climate, etc. See, however, Minton and Malhotra (2007) for a different view. Other recently determined values of $AU_{met}$ are 149,597,870,687.7 ±1.5 m (formal error) (Pitjeva 1997) and 149,597,870,696.0 m ± 0.1 m (Pitjeva 2005). The DE403 ephemeris (Standish et al 1995, IAU 2003 ) uses the primary constant 149,597,870,691 m ± 6 m for the period 1410 BC to 3000 AD. These claimed accuracies are out of line with the variation of $a_\oplus$ in the timespans of the ephemerides. The DE406 ephemeris extends (Standish 2005a) from 3000 BC to 3000 AD with a stated *interpolation* accuracy of 25 m for all planets. The variation in the Earth's semi-major axis from 3000 BC to present, due to $\dot{\mu}$, however, is about 68 - 78 m. With such small quoted errors in the AU and the ephemerides, it is worthwhile to account for $\dot{\mu}$, and to revise the definitions so as to stabilize the AU. The only obvious way to include the effect of changing $M_S$ in post-Newtonian ephemeris modeling (Brumberg 1991, Newhall et al 1983) is to retain the traditional form for the equations, simply including the time dependence in the central force.

### 6.1.2 periods, rates

Pitjeva's (2005a) data span a 1961 - 1995 baseline. She obtains a value for the rate of change of $a_\oplus$ (as reported by Standish 2005b) of 5 m cy$^{-1}$. KB obtain $\dot{a}_\oplus = 15 \pm 4$ m cy$^{-1}$ based on modeling of the ephemerides, and examine possible explanations, finding that systematic error is the most likely. Both these values greatly exceed our predicted result of about 1.4 m cy$^{-1}$. There is no plausible explanation for these disagreements but systematic error. Solar mass loss at a level four to twelve times the value 1.4 m cy$^{-1}$ derived here would require, for example, axion emission 3 to 11 times the photon luminosity, violating the limit of Schlattl et al (1999) by more than an order of magnitude. Cosmological effects can be excluded (Section 7).

A more promising test than verifying Equation (6.1) is to look for a planet's orbital motion falling behind that it would execute at constant solar mass. Integrating Equation (4.4) around successive orbits, with the use of Equations (4.3) and (4.5), yields for the true anomaly

$$\theta = \int \frac{\ell}{r^2(t)} dt \approx \int \frac{\ell}{r_0^2(t)}\left[1 + 2\frac{\dot{\mu}}{\mu}t + \cdots\right] dt \equiv \theta_0 + \Delta\theta \qquad (6.2)$$

which defines $\Delta\theta$ as the *difference* in true anomaly due to solar mass loss; here $r_0$ is the value of $r$ ignoring the change in $\mu$. Specifically,

$$\Delta\ddot{\theta} = \frac{2\ell}{r^2}\frac{\dot{\mu}}{\mu} \qquad (6.3)$$

where we have dropped the subscript on $r$, which indicated zero order. and the double overdots refer to the second time derivative of the quantity $\Delta\theta$. Equation (6.3) is to be applied at corresponding values of $\theta$. With a sufficiently long time base, this effect may be detected as a falling back of the planet's position along its orbit. Using Equation (2.2), and dropping out an assumed $2\pi$ radians per baseline period $P_0$ of the planet, shows that to first order in the lost mass $\Delta\theta$ is *quadratic* in time.[2] In convenient units of the planet's period $P$ and the century,

$$\Delta\ddot{\theta} = -5.7\times10^{-11}(1+0.7442*axratio) \text{ rad P}^{-1}\text{ cy}^{-1}. \qquad (6.4)$$

We apply this to planets Mercury through Uranus and the main belt asteroid (listed here as a "planet") Vesta, for the cases *axratio* = 0.0 and *axratio* = 0.1  The distances along orbit at radius *a* are shown in Table 1.  We refer to the lag "in a century" rather than "per century" because the time dependence is quadratic, not linear. If, instead, we evaluate at aphelion, the changes are larger, mainly for Mercury and Mars (Table 2).

The induced displacements along the orbit, for the four inner planets, from 3000 BC, the start epoch of DE406, to 2008 are 3450 km, 2500 km, 2150 km, and 1740 km, respectively (for no axions, i.e. *axratio* = 0).  Neglecting $\dot{\mu}$ would put Mercury's position ~5000 y ago in error by more than half its diameter, Vesta's by 2.5 times its diameter. The changes in the tables were computed as if the predicted positions were known rigorously for a theory using constant $\mu$; in practice, a theory is fitted to the data, with the effect that the time base is effectively halved, because times are reckoned from the midpoint. With a time half-baseline (midpoint to end) available for radar astronomy of the inner planets ~(46/2) y, displacements would be 73 m, 53 m, 45 m, and 37 m, values that would be significant, as DE406 documentation (Standish 2005a) lists the accuracy for all the planets as "no worse than 25 m, adequate for nearly any application."  But caveats (Standish and Fienga 2002) show that the accuracy deteriorates more than suggested in (Standish 2005a) (see Section 6.2.2). Even at the rate 2 km cy$^{-1}$, most applicable to Mars, Mercury's positional error due to asteroids in 23 y would be 460 m, about ten times the expected effect.  Hence we look forward to refinement in mass estimates for the asteroids.  Fienga et al (2008) have included the planetary and mutual perturbations due to 300 asteroids, but find results similar to recent JPL, Russian and other ephemerides.

Changes in planetary periods,

$$\frac{dP}{P} = 2\frac{da}{a} = -2\frac{dM_\odot}{M_\odot} \qquad (6.5)$$

might be detectable over an interval ~ a century, but historical data do not support the required accuracy, so it is a future research item. Note that the factor 2 in the first equality differs from the usual factor 1.5 because the mass is changing. The usual 1.5 factor applies, of course, to comparison of orbital periods of *different bodies* at *fixed major axes*, not of one body whose major axis changes. The U.S. Naval Observatory Master Clock has deviations from the international average typically 5 – 10 ns, which can be taken as representative of the practical ability to measure time differences over a year or two (Matsakis, 2005). Aside from technical difficulties in measuring absolute periods, the required accuracies for the inner planets are ~ $4 \times 10^{-14} - 6 \times 10^{-13}$ s per period for the inner planets. For the outer planets the nominal period changes in seconds are a hundred times or more larger, but the problem of determining completion of a period is too severe because of the length of time required and the need to use optical positional measurements. Comparing ephemerides with measurements of planetary positions over an extended time interval will always suffer from uncertainties in the time. But *relative* planetary positions could be more definitive, because the changes in position (Tables 1, 2) differ quadratically in time, and so cannot be mimicked by changes in the timescale.

### 6.1.3 occultations

The observational time base can in principle be extended by considering ancient conjunctions and occultations. Chinese records from 146 BCE to 1761 CE have been analyzed by Hilton et al (1988) and compared with the DE102 ephemeris. There was agreement with 300 arc seconds, showing "no serious flaws" with the planetary theories and the rotational deceleration of the Earth. This uncertainty is orders of magnitude greater, however, than what would be desirable for validating the effect of solar mass loss.

## *6.2 Error sources*

### 6.2.1 Could changes in G corrupt our result?

Certain tests for a "changing constant of gravity" are related to the present case. Such changes would affect not only the orbits of all bodies orbiting the sun, but *also* the Moon's rate of recession from Earth. The most stringent direct bound for the rate of change constant of gravity, in fact, comes from analysis of the lunar orbit by Williams et al (2004), who obtain a rate $\dot{G}/G = 4 \pm 9 \times 10^{-13}$ yr$^{-1}$, equivalent to about $\pm 15$ m cy$^{-1}$ change in $a_\oplus$, which could exceed our value ~1.37 – 1.57 m cy$^{-1}$, based on $\dot{M}_\odot/M_\odot$ alone, though it is consistent with zero. The Williams et al result is independent of solar mass changes (within reason). See also Guenther et al (1998) for a helioseismological approach.[3] Benvenuto et al (2004) obtain less stringent limits $-2.5 \times 10^{-10} < \dot{G}/G < 4 \times 10^{-11}$ per year from white dwarf oscillations. Pitjeva (2005b) obtained the stronger limit, $\dot{G}/G = (-2 \pm 5) \times 10^{-14}$ yr$^{-1}$ (applicable to change of the product $\mu = GM_\odot$) from radar observations of planets and spacecraft. This would violate Eq. (2.2) at the 2 σ level, but according to Pitjeva (2008) the tightest limit that can actually be claimed is $|\dot{G}/G| < 10^{-13} \, \text{yr}^{-1}$.

---

[3] Christensen-Dalsgaard et al (2005), however, assert that there are additional sources of error beyond those in Guenther et al.

Given the rather weak limit coming directly from these works, we apply some theoretical analysis to interplanetary data on the spacetime metric. Scalar-tensor gravitational theories, such as Brans-Dicke, (see Will, 1993 for variants) that would tolerate a change in Newtonian $G$ typically give values

$$\dot{G}/G \sim -H_0 * a / (b + \omega_{BD}) \tag{6.6}$$

where $H_0$ is the Hubble parameter, $\omega_{BD}$ the Brans-Dicke coupling constant (or scalar-tensor; the subscript is intended as generic for scalar-tensor) (Weinberg 1972, Will 1993, Faraoni 2004) and $a$ and $b$ are positive numerical constants of order unity, the values $1 < a < 3$ and $b = 2$ being the most common. We adopt the value 2 as exemplary of both parameters.

The value of the constant $\omega_{BD}$ can be obtained from Post-Newtonian fits to the metric in the solar system as measured via the deflection of radio waves by the solar gravity using VLBI (Shapiro et al 2004) or the time delay of radio waves passing near the Sun (Shapiro et al 1971, Bertotti et al 2003). The post-Newtonian parameter directly related to $\omega_{BD}$ is $\gamma$, in the expression $g_{ij} = \delta_{ij}(1 + 2\gamma U)$, where the $g_{ij}$ are the spatial metric coefficients, $\delta_{ij}$ the 3-dimensional Kronecker matrix, and $U$ is the zero-order Newtonian potential (counted positive) (Richter and Matzner 1982). The value of $\gamma$ for the Brans-Dicke theory is (Will 1993)

$$\gamma = \frac{1 + \omega_{BD}}{2 + \omega_{BD}}. \tag{6.7}$$

The most accurate determination of $\gamma$ appears to be that of Bertotti et al (2003)

$$\gamma = 1 + (2.1 \pm 2.3) \times 10^{-5}. \tag{6.8}$$

Kopeikin et al (2006) have, however, asserted that by neglecting the motion of the Sun around the barycenter of the solar system, Bertotti et al (2003) have an accuracy only $\sim \pm 10^{-4}$. Kopeikin et al suggest that re-analysis of the Cassini data could improve that limit. In view of Equation (6.7) and the necessity that $\omega_{BD}$ be positive (Noerdlinger 1968), $\gamma$ must be < 1, so we must take the lower sign in Equation (6.8). This implies $\gamma > 1 - 2 \times 10^{-6}$ or, in turn, $\omega_{BD} > 5 \times 10^5$, accepting Bertotti et al's result, or $\gamma > 1 - 10^{-4}$ if we accept the Kopeikin et al (2006) approach. Using Equation (6.6) with a = b = 2 and the Bertotti result, we find that for $H_0 = 75$ km s$^{-1}$ Mpc$^{-1}$

$$\left| \dot{G}/G \right| \le 4 \times 10^{-6} H_0 < 3 \times 10^{-16} \text{ yr}^{-1}, \tag{6.9}$$

which means that, under the rubric of almost any cosmology presently studied, we can neglect the rate of change of $G$ in that of the product $GM_\odot$. With the maximum Kopeikin et al (2006) error estimate, we would have $\left| \dot{G}/G \right| \le 2 \times 10^{-4} H_0 < 1.5 \times 10^{-14}$ yr$^{-1}$, which is still small (though not entirely negligible) compared with the effect of mass loss. It would be interesting to see a reworking of the Bertotti et al (2003) analysis following the suggestions of Kopeikin et al (2006) since it could reduce this possible source of uncertainty in $\dot{M}_\odot / M_\odot$. The estimates of the rate of change of $a_\oplus$ $M_\odot$ and G according to several authors are summarized in Table 3.

## 6.2.2 The effect of observational error and poorly known asteroid masses

Standish (2004) gives a summary of error estimates for the recent JPL ephemerides. For the inner planets, accurate distance measurements yield errors in DE405 growing at only ~2 km cy$^{-1}$, while those (Table 1) due to ignoring the change in solar mass are comparable. Thus, over a time period ~ 1 cy, the mass loss effect will dominate. For the outer planets, the data are almost all optical positions, and the errors are ~ 0.1 – 1.0 arc seconds, yielding positional errors ~ 500 km (it is not known how these errors would grow with time, because the best data sets are for too short a timespan compared to the orbital periods, so we take the errors as constants.) We expect that the improvement of data with, for example, GAIA observations (Hestroffer and Morando 1995), will outrace any deterioration in the ephemerides due to inaccurate positions and short time spans.

Standish and Fienga (2002) point out that the uncertainties in the masses of asteroids cause steady deterioration of the accuracy of ephemerides. The error in planetary positions grows roughly linearly at about 2.5 km per decade (Bretagnon 2002, Standish 2002), though this could be improved as asteroid masses are modeled better. The quoted value is for Mars (as it is near the asteroid belt, and has resonances with massive asteroids) and would be less for the other planets. The linear behavior of the error can be regarded as an initial ramp to chaotic behavior with a Lyapunov exponent in the millions of years range (Standish 2006). The effect can be separated from the effect of the rate of change in the AU, because the latter grows linearly in the semi-major axis and, tellingly, quadratically in the mean anomaly, while the asteroid mass effect grows linearly in all coordinates, and is greatest for Mars. These errors are largely subsumed in Standish's (2004) discussion.

## 7. Relativistic and cosmological considerations

It is of some merit to pursue these effects in a relativistic analysis, mainly because the calculations for modern solar system ephemerides use post-Newtonian (linearized) relativity. Inclusion of the gravity field of the solar radiation would not appear to add much information, since its equivalent mass is small and slowly varying in time. The metric for a radiating body (Lindquist et al, 1965; Noerdlinger, 1976) has been analyzed in some detail, but no one has succeeded thus far in putting it in time-orthogonal form, and hence also not in Post-Newtonian form. Presumably, the results of such a theory would be close to those obtained by using the Post-Newtonian equations as derived for constant masses, but inserting the varying solar mass. The usual development of the Post-Newtonian formalism (Will, 1993; Soffel et al 2003) uses source equations based on a slowly moving ideal fluid. The stress-energy tensor of the solar radiation field is too different from such a nearly isotropic tensor for one to attempt such a treatment. KB consider possible effects of the expanding universe on the scale of the solar system. In agreement with previous analyses, (Noerdlinger and Petrosian 1971, Cooperstock et al 1998), they find negligible effect. The integrals provided by Soffel et al, as approved in the IAU 2000 Resolutions for dealing with interplanetary gas and radiation appear not to have been used in preparing ephemerides, though some of the terms contribute to frame-dragging effects when solid material is the source. The effects of the cosmological constant $\Lambda$ within the solar system and for binary pulsars have also been estimated by Jetzer and Sereno (2006), who find limits less stringent than those in cosmology by a factor $\sim 10^{10}$. While they find, for acceptable values of $\Lambda$, no effect that would change any of our results, they do find secular changes in the argument of perihelion. In summary, post-Newtonian calculations can safely ignore the expansion of the universe and the cosmological constant.

## 8. Conclusions

Solar mass loss is causing the orbits of the planets to expand linearly, and their angular motions to fall behind quadratically with time. The size of these disturbances is sufficient that long term ephemerides should take the mass loss into account. The mass loss, coupled with the IAU specified definition of the AU, causes the latter to decrease in its value in meters, opposite to the solar system's expansion. This virtually demands redefinition of the AU or the Gaussian gravity constant. It is urgent to improve our estimates of asteroid masses so that all these effects can be accurately tested. It would be desirable if the data of Bertotti et al (2003) were re-analyzed along the lines suggested by Kopeikin et al (2006), so as to better separate possible variations in the Newtonian gravity constant from effects of solar mass loss.

## 9. Acknowledgements

I am indebted to a referee and to Drs. Elena Pitjeva, E. Myles Standish, Kenneth Seidelmann, Sergei Klioner, George Kaplan, E. L. Wright, James Williams, Wayne Hayes and Clifford Will for helpful communications. I am indebted to Dr. A. J. Hundhausen for providing a prepublication copy of his work, and to Dr. O.R. White for calling that work to my attention.

Table 1

Lag along orbit in one century for r = semi-major axis a

| Planet | Lag in orbit (m), no axion loss | Lag in orbit (m), 10% axion loss |
| --- | --- | --- |
| Mercury | 1350 | 1451 |
| Venus | 1009 | 1084 |
| Earth | 858 | 922 |
| Mars | 692 | 744 |
| Vesta | 556 | 598 |
| Jupiter | 376 | 494 |
| Saturn | 278 | 298 |
| Uranus | 196 | 210 |

---

[5] I have taken the liberty of slightly reworking Williams and Standish's material to use SI units of length, rather than km. The distinction between TDB and the ordinary SI (or TAI) (International Atomic Time) time stream is not critical to the present discussion, so long as one or the other time stream is used consistently. For the definition of TDB, see Fukushima, "Time Systems in General Relativity," in Guinot, (1989), Section 4.2.

Table 2

Lag along orbit in one century at aphelion

| Planet | Lag in orbit (m), no axion loss | Lag in orbit (m), 10% axion loss |
|---|---|---|
| Mercury | 1628 | 1749 |
| Venus | 1016 | 1092 |
| Earth | 873 | 938 |
| Mars | 757 | 813 |
| Vesta | 606 | 651 |
| Jupiter | 394 | 423 |
| Saturn | 293 | 314 |
| Uranus | 205 | 220 |

Table 3

Expansion of $a$ and estimates of rates of change of $G$ and $M_\odot$ by various authors

| Author(s) | $\dot{a}\,(\mathrm{m\,cy^{-1}})$ | $\dot{a}/a\,(\mathrm{cy^{-1}})$ | $-\dot{G}/G\,(\mathrm{cy^{-1}})$ | $-\dot{M}_\odot/M_\odot\,(\mathrm{cy^{-1}})$ |
|---|---|---|---|---|
| Krasinsky & Brumberg | $15 \pm 4$ (Eq. 23) or $< 5$ (§4) | $10^{-10}$ | $< 7 \times 10^{-12}$ (Eq. 10) | $3 \times 10^{-12}$ |
| Pitjeva 2005 | 5 | $\sim 3.3 \times 10^{-11}$ | $(2 \pm 5) \times 10^{-12}$ | n/a |
| Pitjeva 2008 | $\sim 1.5$ | $10^{-11}$ | $10^{-11}$ | n/a |
| Williams 2004 | n/a | n/a | $-4 \pm 9 \times 10^{-11}$ | n/a |
| This paper | 1.37 to 1.57 [a] | $> 9.1 \times 10^{-12}$ | $< 1.5 \times 10^{-12}$ [b] | $> 9.1 \times 10^{-12}$ |

---

[a] first number for case of no axions, larger (second) number for maximum axions (per Schlattl et al, 1999)

---

[b] based on the Kopeikin et al (2006) result for the PPN parameter $\gamma$.